\documentclass[11pt]{article}
\usepackage{latexsym}  
\usepackage{amssymb}
\usepackage{graphicx}
\usepackage{amsmath}

\begin{document}

\begin{center}
{\Large\bf Universal Bilinear Form of Quark and Lepton  Mass Matrices} 
  
\vspace{4mm}
{\bf Yoshio Koide$^a$ and Hiroyuki Nishiura$^b$}

${}^a$ {\it Department of Physics, Osaka University, 
Toyonaka, Osaka 560-0043, Japan} \\
{\it E-mail address: koide@kuno-g.phys.sci.osaka-u.ac.jp}

${}^b$ {\it Faculty of Information Science and Technology, 
Osaka Institute of Technology, 
Hirakata, Osaka 573-0196, Japan}\\
{\it E-mail address: nishiura@is.oit.ac.jp}

\date{\today}
\end{center}

\vspace{3mm}

\begin{abstract}
In the so-called ``yukawaon" model, the (effective) Yukawa coupling 
constants $Y_f^{eff}$ are given by vacuum expectation values (VEVs) of 
scalars $Y_f$ (yukawaons) with $3\times 3$ components. 
In the present model, all of the VEV matrices $\langle Y_f \rangle$ 
are given by a bilinear form of VEVs of flavons $\Phi_f$, 
$\langle Y_f \rangle_i^{\ j} = k_f \langle \Phi_f\rangle_{ik} 
\langle \bar{\Phi}_f\rangle^{kj}$, where
$\Phi_f$ is assigned to ${\bf 6}$ of U(3) family symmetry.
As input parameters with family-number dependent values, 
we use only charged lepton mass values.  
Under this formulation, we can give reasonable values of 
quark and lepton masses and their mixings.  
A $CP$ violating phase $\delta_{CP}^\ell=26^\circ$ in the lepton 
sector is predicted. 
The effective Majorana neutrino mass is also predicted.
\end{abstract}

PCAC numbers:  
  11.30.Hv, 
  12.15.Ff, 
  14.60.Pq,  
  12.60.-i, 

\vspace{3mm}

\noindent{\large\bf 1 \ Introduction}

It is an interesting subject in the particle physics 
to investigate whether the observed hierarchical mass spectra and mixings 
of quarks and leptons result from a single origin or not.
In this paper, we try to describe quark and lepton mass matrices
by using only the observed values of charged lepton masses 
$(m_e, m_\mu, m_\tau)$ as input parameters with family-number dependent values, 
and thereby, we investigate 
whether we can describe all other observed mass spectra 
(quark and neutrino mass spectra) and mixings 
(the Cabibbo-Kobayashi-Maskawa \cite{CKM} (CKM) mixing and 
the Pontecorvo-Maki-Nakagawa-Sakata \cite{PMNS} (PMNS) mixing) without 
using any other family-number dependent parameters. 
Here, terminology ``family-number independent parameters" means, 
for example, coefficients of a unit matrix ${\bf 1}$, 
a democratic matrix $X_3$, and so on, where
$$
{\bf 1} = \left( 
\begin{array}{ccc}
1 & 0 & 0 \\
0 & 1 & 0 \\
0 & 0 & 1 
\end{array} \right) , \ \ \ \ \ 
X_3 = \frac{1}{3} \left( 
\begin{array}{ccc}
1 & 1 & 1 \\
1 & 1 & 1 \\
1 & 1 & 1 
\end{array} \right) . 
\eqno(1.1)
$$
On the other hand, terminology ``family-number dependent parameter" means, 
for example, coefficients of 
$$
{\bf 1}_3 = \left( 
\begin{array}{ccc}
0 & 0 & 0 \\
0 & 0 & 0 \\
0 & 0 & 1 
\end{array} \right) , \ \ \ \ \ 
X_2 = \frac{1}{2} \left( 
\begin{array}{ccc}
1 & 1 & 0 \\
1 & 1 & 0 \\
0 & 0 & 0 
\end{array} \right) . 
\eqno(1.2)
$$

For such our purpose, in this paper, the investigation is done 
on the bases of the so-called yukawaon model
\cite{yukawaon_models,K-N_PRD13}.
Here, the (effective) Yukawa coupling constants $Y_f^{eff}$ are given 
by vacuum expectation values (VEVs) of 
scalars $Y_f$ (yukawaons) with $3\times 3$ components
$$
(Y_f^{eff})_i^{\ j} = \frac{y_f}{\Lambda} \langle (Y_f)_i^{\ j}\rangle 
\ \ \ \ (f=u, d, \nu, e),
\eqno(1.3)
$$
where $\Lambda$ is a scale of the effective theory.
The conception of ``yukawaons" are summarized as follows:
(i) Yukawaons are a kind of flavons \cite{flavon}.
(ii) Those are singlets under the conventional gauge symmetries.
(iii) Since yukawaons are fields, we can consider a non-Abelian
family symmetry $G$ by assigning suitable quantum numbers to $Y_f$. 
(In the present paper, we will assume G=U(3).)
(iv) The VEV forms are described by $3\times 3$ matrices. 
(v) Each yukawaon is distinguished from others by $R$ charges. 
(vi) VEV matrix relations are calculated from SUSY vacuum 
conditions.  
The relations are given by multiplicative forms among VEV matrices
(e.g. $M_R = M_u^{1/2} M_e + M_e M_u^{1/2}$, and so on),
differently from the conventional family symmetry models, 
in which mass matrix form is given by forms of additions  
(e.g. $M= c_1 M_1 + c_2 M_2 + \cdots$). 
(vii) The VEV matrix $\langle Y_f \rangle$ also evolves after the 
family symmetry breaking in the same way that a conventional 
Yukawa coupling constant in the standard model (SM) evolves.

\begin{table}
\caption{Contrast of VEV relations in present yukawaon model to those 
in the previous yukawaon model \cite{K-N_PRD13}.  
For simplicity, notations ``$\langle$" and ``$\rangle$" are drop. }
\begin{center}
\begin{tabular}{ll} \hline
Previous &  Present \\ \hline
$Y_e = k_e \Phi_0 ( {\bf 1} + a_e X_3) \Phi_0$  , &  
$Y_e = k_e \Phi_e \Phi_e$ ,  \\ 
              &  $\Phi_e = k'_e \Phi_0 ( {\bf 1} + a_e X_3) \Phi_0$ , \\
$Y_\nu = k_\nu \Phi_0 ( {\bf 1} + a_\nu X_2) \Phi_0$  , &
$Y_\nu = k_\nu \Phi_\nu \Phi_\nu + \xi_\nu {\bf 1}$ ,  \\ 
            &  $\Phi_\nu = k'_\nu \Phi_0 ( {\bf 1} + a_\nu X_3) \Phi_0$ , \\
$Y_u = k_u P_u \Phi_u \Phi_u P_u^\dagger$, & 
$Y_u = k_u \Phi_u \Phi_u + \xi_u {\bf 1}$, \\
$\Phi_u = k'_u \Phi_0 ( {\bf 1} + a_u X_3) \Phi_0$ , & 
$P_u \Phi_u P_u = k'_u \Phi_0 ( {\bf 1} + a_u X_3) \Phi_0$ , \\
$Y_d = k_d \Phi_d \Phi_d $, & $Y_d = k_d \Phi_d \Phi_d$, \\ 
$\Phi_d = k'_d \Phi_0 ( {\bf 1} + a_d X_3) \Phi_0 + \xi_d {\bf 1}$ , &
$\Phi_d  = k'_d \Phi_0 ( {\bf 1} + a_d X_3) \Phi_0 + \xi'_d {\bf 1}$ , \\ 
$M_\nu =[Y_\nu Y_R^{-1} Y_\nu]^2$, &  $M_\nu =Y_\nu Y_R^{-1} Y_\nu$, \\
$Y_R =  Y_e \Phi_u + \Phi_u Y_e$, &  $Y_R =  Y_e \Phi_u + \Phi_u Y_e$, \\
\hline
\end{tabular}
\end{center}
\end{table}

In order to see differences between the new model and the previous 
yukawaon model \cite{K-N_PRD13}, 
we have listed the VEV relations  of flavons in the present model  
in comparison to those in the previous yukawaon model in Table 1. 
Here, the VEV matrices $Y_e$, $Y_\nu$, $Y_u$ and $Y_d$ correspond to 
charged lepton mass matrix $M_e$, neutrino Dirac mass matrix $M_D$, 
up-quark mass matrix $M_u$, and down-quark mass matrix $M_d$, respectively. 
For simplicity, we have dropped family indices although we consider 
family symmetries U(3)$\times$U(3)$'$. Also, notations ``$\langle$" 
and ``$\rangle$" were drop for simplicity.

\vspace{2mm}

\noindent{\bf  \ VEV relations  of flavons in the previous yukawaon model}

Prior to describing of a new yukawaon model, let us give a brief review 
of the previous yukawaon model \cite{K-N_PRD13}.
The essential VEV relations of flavons in the previous yukawaon model are
listed in the left row in Table 1.
As seen in Table 1, the previous yukawaon model has the following characteristics:  
(i) When we regard the form $\Phi_0 ({\bf 1} +a_f X_3) \Phi_0$ 
($\Phi_0$ is a diagonal VEV matrix) as one unit, $Y_u$ and $Y_d$ take
bilinear forms, while $Y_e$ and $Y_\nu$ are not so.
(ii) Since $a_e \neq 0$, the VEV matix $Y_e$ is not diagonal.  
In an earlier version \cite{yukawaon-K-mass} of the yukawaon model,  
the VEV matrix $Y_e$ was given by a bilinear form $Y_e=\Phi_e \Phi_e$
($\Phi_e$ corresponds to $\Phi_0$ in the previous model \cite{K-N_PRD13}), 
and thereby, a charged lepton mass relation \cite{Koidemass}
$$
K= \frac{m_e + m_\mu + m_\tau}{\sqrt{m_e} + \sqrt{m_\mu} +
\sqrt{m_\tau})^2} = \frac{2}{3} ,
\eqno(1.4)
$$
was speculated by considering a VEV matrix relation 
${\rm Tr}[\Phi_e\Phi_e] = \frac{2}{3} {\rm Tr}[\Phi_e] {\rm Tr}[\Phi_e]$.
However, since the VEV natrux $Y_e$ in the previous model do not take 
such a bilinear form, we cannot speculate the mass relation (1.4). 
The explanation of the formula (1.4) was one of the motivations of 
the yukawaon model.
(iii) The matrix $Y_\nu$ contains the family-number dependent VEV 
matrix form $X_2$ which is defined in Eq.(1.2).  
The VEV matrix $X_2$ was brought in the model together with the
unwelcome condition $a_e \neq 0$  in order to give 
the observed large neutrino mixing 
$\sin^2 2\theta_{13}\simeq 0.09$ \cite{theta13}. 
However, our goal is a model without such a VEV matrix $X_2$.
(iv) Neutrino mass matrix $M_\nu$ was given by a double seesaw 
form (the so-called ``inverse seesaw" form \cite{inverse_seesaw}) 
$M_\nu =(Y_\nu Y_R^{-1} Y_\nu)^2$, where $Y_\nu$ and $Y_R$
are Dirac and Majorana neutriono mass matrices, respectively.
The form was requested in order to give a reasonable ratio 
of neutrino squared mass difference, $R_\nu$, which is defined in 
Eq.(3.14) later.  

\vspace{2mm}

\noindent{\bf  \ VEV relations  of flavons in the present yukawaon model}

The essential VEV relations of flavons in the present yukawaon model are
listed in the right row in Table 1. 
The new model has the following characteristics:  
(i) VEV matrices of all yukawaons have the same family structure, while, 
in the previous yukawaon model, those were taken
different forms for individual sectors.  
(ii) In the previous model, $Y_e$ was not diagonal.
However, in the new model, we succeed in building a model with
$a_e=0$, i.e. a charged lepton mass matrix with a diagonal 
form. 
In the new model, the VEV matrix $\Phi_e$ is diagonal, and
given by 
$$
\Phi_e = k'_e {\rm diag} (m_e^{1/2}, m_\mu^{1/2}, m_\tau^{1/2}).
\eqno(1.5)
$$
Therefore, we again has a possibility that 
the model leads to 
a charged lepton mass relation (1.4).
(However, in this paper, we do not discuss the details.)
(iii) In the previous model, in order to give a large value 
of lepton mixing parameter $\sin^2 2\theta_{13}\simeq 0.09$, 
we were obligated to bring an unwelcome VEV form $Y_\nu$, 
i.e. a family-number dependent form
$Y_\nu = \Phi_0 ( {\bf 1} + a_\nu X_2 ) \Phi_0$.
In contrast to the previous model,  
the present model has succeeded in removing such the
family-number dependent VEV matrix form $X_2$, and 
in unifying VEV matrix forms $\Phi_f$ into the form 
$\Phi_f = \Phi_0 ( {\bf 1} + a_f X_3) \Phi_0$.  
(iv) Neutrino mass matrix is again simply taken as
$M_\nu = Y_\nu Y_R^{-1} Y_\nu$ differently from 
$M_\nu = Y_\nu Y_R^{-1} Y_\nu \cdot Y_\nu Y_R^{-1} Y_\nu$  
in the previous model.

We would like to emphasize that the purpose of the yukawaon model 
is to build a unified mass matrix model of quarks and leptons 
without introducing family-dependent parameters 
(as few as possible) except for the input values $(m_e, m_\mu, m_\tau)$. 
It is not our main purpose to build a model with economized parameters.
Differently from  conventional mass matrix model  with a universal form 
(for a recent model, see, for example, Ref.\cite{Gu14}), we do not adhere to 
a universal form of mass matrices. 
In this paper, we propose a universal bilinear form  of quark  and 
lepton mass matrices.  
However, it is a by-product of our purpose, and our purpose itself 
is not to obtain a universal form of mass matrices.

In Sec.2, we will give details of the VEV matrix 
relations and superpotentials which give such VEV relations. 
In the yukawaon model, $R$ charge assignments are essential for 
obtaining successful phenomenological results. 
Although we assign $R$ charges from the phenomenological 
point of view, the assignments cannot be taken freely. 
We must take the assignments so that they may forbid  
appearance of unwelcome terms. 
The details are also discussed in Sec.2.
In Sec.3, we give a parameter fitting under the new yukawaon model.
Finally Sec.4 is devoted to a summary and concluding remarks.


\vspace{2mm}

\noindent{\large\bf 2 \ Superpotential and VEV matrix relations }

We assume that a would-be Yukawa interaction which is invariant under 
a family symmetry U(3) is given as follows:
$$
W_Y = \frac{y_\nu}{\Lambda} (\nu^c)^i (\hat{Y}_\nu^T)_i^{\ j} \ell_j  H_u 
+ \frac{y_e}{\Lambda} (e^c)^i (\hat{Y}_e)_i^{\ j}\ell_j  H_d 
+ y_R (\nu^c)^i (Y_R)_{ij} (\nu^c)^j 
$$
$$
+ \frac{y_u}{\Lambda}  (u^c)^i (\hat{Y}_u)_i^{\ j} q_j H_u 
+ \frac{y_d}{\Lambda}  (d^c)^i (\hat{Y}_d)_i^{\ j} q_{j} H_d  ,
\eqno(2.1)
$$
where $\ell=(\nu_L, e_L)$ and $q=(u_L, d_L)$ are SU(2)$_L$ doublets.  
The third term in Eq.(2.1) leads to the so-called neutrino seesaw mass 
matrix \cite{seesaw} 
$M_\nu =\hat{Y}_\nu Y_R^{-1} \hat{Y}^T_\nu$, where $\hat{Y}_\nu$ and $Y_R$ 
correspond to neutrino Dirac and Majorana mass matrices, respectively.  
Here and hereafter, for convenience, we use notation 
$\hat{A}$, $A$ and $\bar{A}$ for fields with ${\bf 8}+{\bf 1}$,
${\bf 6}$ and ${\bf 6}^*$ of U(3), respectively.

In order to distinguish each yukawaon from others, we assume that
$\hat{Y}_f$ have different $R$ charges from each other together 
with considering 
$R$ charge conservation (a global U(1) symmetry in $N=1$ supersymmetry).
(Of course, the $R$ charge conservation is broken
at an energy scale $\Lambda$, at which the U(3) family symmetry 
is broken.)
For $R$ parity assignments, we inherit those 
in the standard SUSY model, so 
$R$ parities of yukawaons $Y_f$ (and all flavons) are the same 
as those of Higgs particles 
(i.e. $P_R({\rm fermion})=-1$ and $P_R({\rm scalar})=+1$), 
while quarks and leptons are assigned to 
$P_R({\rm fermion})=+1$ and $P_R({\rm scalar})=-1$.

VEV relations among those yukawaons are obtained from SUSY vacuum conditions for  
superpotentials as we give later.
Here, we need to introduce  
subsidiary flavons which have  special VEV forms:
$$
\langle E \rangle ={\bf 1}, \ \ \langle \bar{E} \rangle ={\bf 1}, 
\eqno(2.2)
$$
$$
\langle P_u \rangle = {\rm diag}(e^{i\phi_1},\ e^{i\phi_2},\ 1) , \ \ 
\langle \bar{P}_u \rangle = {\rm diag}(e^{-i\phi_1},\ e^{-i\phi_2},\ 1) , \ \
\eqno(2.3)
$$ 
$$
\langle \Phi_0 \rangle = {\rm diag}(x_1, x_2, x_3), \ \ \ 
\langle \bar{\Phi}_0 \rangle = {\rm diag}(x_1, x_2, x_3),
\eqno(2.4)
$$
$$
\langle S_f\rangle = \left( {\bf 1} 
+ a_f {X}_3 \right), \ \ \
\langle\bar{S}_f\rangle =\left( {\bf 1}
+ a_f {X}_3 \right),
\eqno(2.5)
$$
where we have dropped  flavor-independent factors in those VEV matrices, 
because we deal with only mass ratios and mixings in this paper.
The forms (2.4) and (2.5) are discussed later. 
(In (2.4) and (2.5), we have introduced another symmetry U(3)$'$ 
in addition to the U(3) flavor symmetry.)


\vspace{2mm}

{\bf 2.1 \ VEV forms of flavons $E$, $\bar{E}$, $P_u$, and $\bar{P}_u$}


For flavons $E$ and $\bar{E}$, we consider the following superpotential: 
$$
W_{E} = \lambda_{1E} {\rm Tr}[E \bar{E} E \bar{E}] +
\lambda_{2E} {\rm Tr}[E \bar{E}] {\rm Tr}[ E \bar{E}],
\eqno(2.6)
$$
where we have taken $R$ charges such that 
$$
R(E)+R(\bar{E}) = 1.
\eqno(2.7)
$$
The SUSY vacuum condition leads to
$$
  \langle {E} \rangle \langle \bar{E} \rangle = {\bf 1} . 
\eqno(2.8)
$$ 
We choose a special solution of Eq.(2.8), 
$$
\langle E \rangle = \langle \bar{E} \rangle = {\bf 1} .
\eqno(2.9)
$$

For $P_u$ and $\bar{P}_u$, we also consider the 
following superpotential form
$$
W_{P} = \frac{\lambda_{1P}}{\Lambda} {\rm Tr}[P_u \bar{P}_u  P_u \bar{P}_u] +
\frac{\lambda_{2P}}{\Lambda} {\rm Tr}[P_u \bar{P}_u ] {\rm Tr}[ P_u \bar{P}_u],
\eqno(2.10)
$$
where we have taken $R$ charges as
$$
R(P_u)+R(\bar{P}_u) = 1.
\eqno(2.11)
$$
The SUSY vacuum condition leads to
$$
  \langle {P}_u \rangle \langle \bar{P}_u \rangle = {\bf 1} . 
\eqno(2.12)
$$ 
In general, it should be noted that for VEV matrices $\langle A\rangle$ and 
$\langle \bar{A} \rangle$ under the $D$-term condition, 
we can choose either one in two cases
$$
\langle \bar{A} \rangle = \langle {A} \rangle^{*} ,
\eqno(2.13)
$$
$$
\langle \bar{A} \rangle = \langle {A} \rangle .
\eqno(2.14)
$$
We apply the case (2.13) to the VEV matrices  
$\langle P_u\rangle$ and $\langle \bar{P}_u \rangle$. 
Then, we obtain (2.3).


\vspace{2mm}

{\bf 2.2 \ Superpotential forms of yukawaons $\hat{Y}_f$ 
and sub-yukawaons $\Phi_f$}

Let us consider a superpotential for $\hat{Y}_f$  ($f=\nu, e, u, d$),   

$$
W_{\hat{Y}} =  \sum_{f=\nu,e, u, d} \left[ \left( \mu_f (\hat{Y}_f)_i^{\ j}   
+ \lambda_{f} ( \Phi_f)_{ik}(\bar{\Phi}_f)^{kj}  \right)
 (\hat{\Theta}_f)_j^{\ i} 
+ \left( \mu'_f (\hat{Y}_f)_i^{\ i}   
+ \lambda'_{f} ( \Phi_f)_{ik}(\bar{\Phi}_f)^{ki}  \right)
 (\hat{\Theta}_f)_j^{\ j} \right].
\eqno(2.15)
$$
Then, a SUSY vacuum condition $\partial W_{\hat{Y}}/\partial \hat{\Theta}_f=0$ 
leads to VEV relation
$$
\langle \hat{Y}_f \rangle = \langle \Phi_f\rangle \langle\bar{\Phi}_f\rangle 
+\xi_f {\bf 1} , 
\eqno(2.16)
$$
where $\xi_f= {\rm Tr} \left[[\langle\hat{Y}_f\rangle 
+ \langle\Phi_f\rangle \langle\bar{\Phi}_f]\rangle \right]$.
Here and hereafter, according to conventional yukawaon models, 
we have assume that all VEV matrices of the $\Theta$ flavons take
$\langle\Theta \rangle =0$. 
Therefore, SUSY vacuum conditions for other flavons do not 
bring any additional VEV relation. 

Note that the appearance of $\xi_f {\bf 1}$ terms in Eq.(2.16) 
is peculiar to the $\hat{\Theta}$ fields. 
If $\Theta$ fields have been ${\bf 6}$ or ${\bf 6}^*$ of U(3), 
such a $\xi_f{\bf 1}$ would not be able to appear. 
Meanwhile, as shown in Table 1, we have taken $\xi_e =\xi_d=0$.
The reason is purely based on a phenomenological requirement.
(See the next section.) 

For $\Phi_e$ and $\Phi_\nu$, we assume a superpotential
$$
W_{\Phi e, \Phi \nu} = \sum_{f=e, \nu} 
\left( \mu_f (\Phi_f)_{ij} +\lambda_{f} (\Phi_0)_{i\alpha}
(\bar{S}_f)^{\alpha\beta} (\Phi_0^T)_{\beta j} \right) (\bar{\Theta}_f)^{ji},
\eqno(2.17)
$$
which lead to 
$$
\langle \Phi_f \rangle = \langle \Phi_0 \rangle \langle \bar{S}_f \rangle
\langle \Phi_0^T \rangle \ \ \ (f=e, \nu) ,
\eqno(2.18)
$$
where $\Phi_0$ and $S_f$ are new flavons which belong to $({\bf 3},{\bf 3})$ and 
$({\bf 1},{\bf 6}^*)$ of U(3)$\times$U(3)$'$, respectively.
The VEV form of $\Phi_0$ is given by Eq.(2.4). 
In general, we can choose the flavor basis such that $\langle \Phi_0\rangle$ 
is diagonal.  
As we discuss later, since we take $a_e=0$, we can denote Eq.(2.4) as
$$
\langle \Phi_0 \rangle = \langle \bar{\Phi}_0 \rangle = 
{\rm diag} (x_1, x_2, x_3) = 
{\rm diag} (m_e^{1/4}, m_\mu^{1/4}, m_\tau^{1/4}),
\eqno(2.19)
$$
from the $D$-term condition, where $x_i$ are real and
those are normalized as $x_1^2+x_2^2+x_3^2 =1$.
The VEV form of $S_f$ is given by Eq.(2.5). 
We consider that the form (2.5) is due to a symmetry breaking 
U(3)$' \rightarrow$ S$_3$ at $\mu=\Lambda'$.  
(Of course, we assume a superpotential similar to (2.17) 
for the flavons $\bar{\Phi}_f$.

On the other hand, for $\Phi_u$, we assume a form
$$
W_{\Phi u} = \frac{1}{\Lambda} \left( \lambda_{1u} 
(\bar{P}_u)^{ik} (\Phi_u)_{kl} (\bar{P}_u)^{lj} 
+ \lambda_{2u} (\bar{\Phi}_0)^{ik} (S_u)_{kl}  (\bar{\Phi}_0^T)^{lj} \right) 
(\Theta_u)_{ji} ,
\eqno(2.20)
$$
which leads to
$$
\langle \bar{P}_u \rangle  \langle \Phi_u \rangle \langle \bar{P}_u \rangle =
 \langle \bar{\Phi}_0 \rangle \langle S_u \rangle 
\langle \bar{\Phi}_0^T \rangle .
\eqno(2.21)
$$

In order to obtain $\xi'_d {\bf 1}$ term for $\Phi_d$ as shown in Table 1, 
we assume the following superpotential 
$$
W_{\Phi d} = \frac{\lambda_{1d}}{\Lambda} 
{\rm Tr}[\bar{E} \Phi_d \bar{E} \Theta_d ]
+ \frac{\lambda_{2d}}{\Lambda} {\rm Tr}[\bar{\Phi}_0 
S_d \bar{\Phi}_0^T  \Theta_d ]
+  \frac{\lambda_{3d}}{\Lambda} {\rm Tr}
[\bar{E} \Phi_d] {\rm Tr}[\bar{E} \Theta_d ] ,
\eqno(2.22)
$$
which leads to
$$
\langle \bar{E}\rangle \langle \Phi_d \rangle \langle  \bar{E}\rangle =
\langle  \bar{\Phi}_0\rangle \langle S_d\rangle 
\langle \bar{\Phi}_0^T\rangle
+ \xi'_d \langle \bar{E}\rangle ,
\eqno(2.23)
$$
where $\xi'_d =(\lambda_{3d}/\lambda_{1d}){\rm Tr}[\langle \bar{E}
\rangle \langle \Phi_d  \rangle ]$. 
We can also consider a superpotential for $\bar{\Phi}_d$ 
accompanied with $\xi'_d {\bf 1}$. 

Note that in Eq.(2.22) we have added the $\lambda_{3d}$ term to the 
$\lambda_{1d}$ and $\lambda_{2d}$ terms which correspond to 
the $\lambda_{1u}$ and $\lambda_{2u}$ terms in the 
superpotential $W_{\Phi u}$, Eq.(2.20).
If we have considered a $\lambda_{3u}$ term in $W_{\Phi u}$
as well as the $\lambda_{3d}$ term in $W_{\Phi d}$, 
we would obtain
$\langle \bar{P}_u \rangle  \langle \Phi_u \rangle \langle \bar{P}_u \rangle =
 \langle \bar{\Phi}_0 \rangle \langle S_u \rangle 
\langle \bar{\Phi}_0 \rangle+ \xi'_u  \langle \bar{P}_u \rangle$ 
with a complex coefficient  
$\xi'_u \propto {\rm Tr} [ \langle \bar{P}_u \rangle  \langle \Phi_u \rangle]$ 
instead of Eq.(2.21).
Then, not only the CKM parameters, but also the up-quark mass ratios and the
PMNS parameters become dependent on the phase parameters $(\phi_1, \phi_2)$.
We assume that the contribution from the $\lambda_{3u}$ term is negligibly
small from the practical reason for parameter fitting in the next section.

For $Y_R$, we assume a superpotential form 
$$
W_R = \left[ \mu_R (Y_R)_{ij} + \lambda_R \left( (\hat{Y}_e)_i^{\ k} (\Phi_u)_{kj}
+ (\Phi_u)_{ik} (\hat{Y}_e^T)^k_{\ j} \right) \right] (\bar{\Theta}_R)^{ji} ,
\eqno(2.24)
$$
which reads to 
$$
\langle Y_R \rangle = \langle \hat{Y}_e\rangle \langle \Phi_u\rangle
+ \langle \Phi_u \rangle \langle \hat{Y}_e^T\rangle .
\eqno(2.25)
$$ 

The VEV relations described above have been derived dependently  
on the assignments of $R$ charges for the flavons.
The $R$ charge assignments are discussed in the next subsection. 
In the meanwhile, we list the assignments of 
SU(2)$_L \times$SU(3)$_c \times$U(3)$\times$U(3)$'$ 
for the fields which appear in the present model in Table 2. 
As seen in Table 2, the existence number of fields 
with ${\bf 3}$ and ${\bf 3}^*$ 
(and also ${\bf 6}$ and ${\bf 6}^*$) of U(3)-family (and also 
U(3)$'$) are the same, so that the model are anomaly free.

\begin{table}
\caption{Assignments of SU(2)$_L \times$SU(3)$_c \times$U(3)$\times$U(3)$'$. 
For $R$ charges, see subsection 2.3. 
We assign the same $R$ charges   
for flavons $A$ and $\bar{A}$ which have the same VEV 
$\langle A \rangle= \langle \bar{A} \rangle$, 
 e.g. $R(A)=R(\bar{A})$. 
However, since $R(P_u) \neq R(\bar{P}_u)$ because of 
$\langle P_u \rangle \neq \langle \bar{P}_u \rangle$, we have 
$R(S_u) \neq R(\bar{S}_u)$ and 
$R(\Theta_u) \neq R(\bar{\Theta}_u)$, i.e. 
$r_{Su} = 2 \bar{r}_P +r_u -2 r_0$,  $\bar{r}_{Su} = 2 {r}_P +r_u -2 r_0$,
$r_{\Theta u} =2- 2 r_0  -r_{Su}$, and 
$\bar{r}_{\Theta u} =2- 2 r_0  -\bar{r}_{Su}$. 
}

\begin{center}
\begin{tabular}{|c|ccc|ccc|cc|cccc|} \hline
& $\ell$ & $e^c$ & $\nu^c$ & $q$ & $u^c$ & $d^c$ & 
$H_u$ & $H_d$ &  
$\hat{Y}_e$ & $\hat{Y}_\nu$ & $\hat{Y}_u$ & $\hat{Y}_d$ 
\\ \hline
SU(2)$_L$ & ${\bf 2}$ & ${\bf 1}$ & ${\bf 1}$ & 
${\bf 2}$ & ${\bf 1}$ & ${\bf 1}$ & 
${\bf 2}$ & ${\bf 2}$ &
 ${\bf 1}$ & ${\bf 1}$ & ${\bf 1}$ & ${\bf 1}$
\\ 
SU(3)$_c$ & ${\bf 1}$ & ${\bf 1}$ & ${\bf 1}$ & 
${\bf 3}$ & ${\bf 3}^*$ & ${\bf 3}^*$ & ${\bf 1}$ & ${\bf 1}$ &
 ${\bf 1}$ & ${\bf 1}$ & ${\bf 1}$ & ${\bf 1}$ \\ \hline
U(3) & ${\bf 3}$ & ${\bf 3}^*$ & ${\bf 3}^*$ &  
${\bf 3}$ & ${\bf 3}^*$ & ${\bf 3}^*$ & ${\bf 1}$ & ${\bf 1}$ & 
 ${\bf 8}+{\bf 1}$ & ${\bf 8}+{\bf 1}$ & 
 ${\bf 8}+{\bf 1}$ &  ${\bf 8}+{\bf 1}$ \\
U(3)$'$ & ${\bf 1}$ & ${\bf 1}$ & ${\bf 1}$ &  
${\bf 1}$ & ${\bf 1}$ & ${\bf 1}$ & ${\bf 1}$ & ${\bf 1}$ & 
${\bf 1}$ & ${\bf 1}$ & ${\bf 1}$ & ${\bf 1}$ 
\\  \hline
$R$     & 2 & $-2r_{e}$  & $-2r_{\nu}$ & 
2 & $-2r_{u}$  & $-2r_{d}$ & 0  & 0 &
$2r_e$ & $2r_\nu$ & $2r_u$ & $2r_d$ 
\\  \hline 
\end{tabular}

\vspace{2mm}

\begin{tabular}{|c|cccccccc|cc|cc|cc|} \hline
 ${Y}_R$ & 
$\bar{\Phi}_e$ & $\Phi_e$ & $\bar{\Phi}_\nu$ & $\Phi_\nu$ &
$\bar{\Phi}_u$ & $\Phi_u$ & $\bar{\Phi}_d$ & $\Phi_d$ & $\bar{P}_u$ &  ${P}_u$ &
 ${\Phi}_0$ & $\bar{\Phi}_0$ 
\\ \hline
 ${\bf 1}$ & ${\bf 1}$ & ${\bf 1}$ &  ${\bf 1}$ & 
${\bf 1}$ & ${\bf 1}$ & ${\bf 1}$ & ${\bf 1}$ & 
${\bf 1}$ & ${\bf 1}$ & ${\bf 1}$ & ${\bf 1}$ & ${\bf 1}$
\\ 
 ${\bf 1}$ & ${\bf 1}$ & ${\bf 1}$ &  ${\bf 1}$ & 
${\bf 1}$ & ${\bf 1}$ & ${\bf 1}$ & ${\bf 1}$ & 
${\bf 1}$ & ${\bf 1}$ & ${\bf 1}$ & ${\bf 1}$ & ${\bf 1}$
\\ \hline
 ${\bf 6}$ &  ${\bf 6}^*$ & ${\bf 6}$ &  
${\bf 6}^*$ & ${\bf 6}$ & ${\bf 6}^*$ & ${\bf 6}$ & 
${\bf 6}^*$ & ${\bf 6}$ & ${\bf 6}^*$ & ${\bf 6}$ 
& ${\bf 3}$ & ${\bf 3}^*$ \\ 
${\bf 1}$ &
${\bf 1}$ & ${\bf 1}$ & ${\bf 1}$ & ${\bf 1}$ & 
${\bf 1}$ & ${\bf 1}$ & ${\bf 1}$ & ${\bf 1}$ & 
${\bf 1}$ & ${\bf 1}$ & ${\bf 3}^*$ & ${\bf 3}$ 
\\ \hline
$r_R$ & 
$r_e$ & $r_e$ & $r_\nu$ & $r_\nu$ & 
$r_u$ & $r_u$ & $r_d$ & $r_d$ & $1-{r}_P$ & $r_P$ &
$r_0$ & $r_0$ 
\\ \hline
\end{tabular}

\vspace{2mm}

\begin{tabular}{|cccccccc|} \hline
$S_e$ & $\bar{S}_e$ & $S_\nu$ & $\bar{S}_\nu$ & 
$S_u$ & $\bar{S}_u$ & $S_d$ & $\bar{S}_d$   
\\ \hline
 ${\bf 1}$ &   ${\bf 1}$ &  ${\bf 1}$ &  ${\bf 1}$ & 
 ${\bf 1}$ &  ${\bf 1}$ &  ${\bf 1}$ & ${\bf 1}$ 
\\ 
 ${\bf 1}$ &   ${\bf 1}$ &  ${\bf 1}$ &  ${\bf 1}$ & 
 ${\bf 1}$ &  ${\bf 1}$ &  ${\bf 1}$ & ${\bf 1}$ 
\\ \hline
 ${\bf 1}$ &  ${\bf 1}$ &  ${\bf 1}$ & ${\bf 1}$ & 
 ${\bf 1}$ &  ${\bf 1}$ &  ${\bf 1}$ & ${\bf 1}$  \\ 
${\bf 6}$ &  ${\bf 6}^*$ &  ${\bf 6}$ & ${\bf 6}^*$ & 
${\bf 6}$ &  ${\bf 6}^*$ &  ${\bf 6}$ & ${\bf 6}^*$  
\\ \hline
\multicolumn{2}{|c}{$r_e -2 r_0$} & 
\multicolumn{2}{c}{$r_\nu -2 r_0$} & $r_{Su}$ & $\bar{r}_{Su}$ & 
\multicolumn{2}{c|}{$r_d +1-2 r_0$}
\\ \hline
\end{tabular}

\vspace{2mm}

\begin{tabular}{|cc|cccc|c|} \hline
 ${E}$ & $\bar{E}$ &
$\hat{\Theta}_e$ & $\hat{\Theta}_\nu$ & $\hat{\Theta}_u$ & $\hat{\Theta}_d$ & 
$\bar{\Theta}_R$ 
\\ \hline
  ${\bf 1}$ &  ${\bf 1}$ & ${\bf 1}$ & ${\bf 1}$ & ${\bf 1}$ &  ${\bf 1}$ & 
 ${\bf 1}$ 
\\ 
 ${\bf 1}$ & ${\bf 1}$ &  ${\bf 1}$ & ${\bf 1}$ & ${\bf 1}$ &  ${\bf 1}$ & 
${\bf 1}$ 
\\ \hline
 ${\bf 6}$ & ${\bf 6}^*$ & ${\bf 8}+{\bf 1}$ &  ${\bf 8}+{\bf 1}$ &  
${\bf 8}+{\bf 1}$ &  ${\bf 8}+{\bf 1}$ & ${\bf 6}^*$ 
\\
${\bf 1}$ &  ${\bf 1}$ &  ${\bf 1}$ & ${\bf 1}$ & ${\bf 1}$ &  ${\bf 1}$ & 
${\bf 1}$ 
\\ \hline
 $\frac{1}{2}$ & $\frac{1}{2}$ &
 $2-2r_e$   & $2-2r_\nu$  & $2-2r_u$  & $2-2r_d$ & $2-2r_R$        
\\ \hline
\end{tabular}

\vspace{2mm}

\begin{tabular}{|cccccccc|} \hline
$\Theta_{\Phi e}$ & $\bar{\Theta}_{\Phi e}$ & 
 $\Theta_{\Phi \nu}$ & $\bar{\Theta}_{\Phi \nu}$ & 
$\Theta_{\Phi u}$ & $\bar{\Theta}_{\Phi u}$ & 
$\Theta_{\Phi d}$ &  $\bar{\Theta}_{\Phi d}$ 
\\ \hline
${\bf 1}$ & ${\bf 1}$ & ${\bf 1}$ &  ${\bf 1}$ & 
${\bf 1}$ &  ${\bf 1}$ &  ${\bf 1}$ &  ${\bf 1}$ 
\\ 
 ${\bf 1}$ & ${\bf 1}$ & ${\bf 1}$ &  ${\bf 1}$ & 
${\bf 1}$ &  ${\bf 1}$ &  ${\bf 1}$ &  ${\bf 1}$ 
\\ \hline
 ${\bf 6}$ & ${\bf 6}^*$ & ${\bf 6}$ &  ${\bf 6}^*$ & 
${\bf 6}$ &  ${\bf 6}^*$ &  ${\bf 6}$ &  ${\bf 6}^*$ 
\\
 ${\bf 1}$ & ${\bf 1}$ & ${\bf 1}$ &  ${\bf 1}$ & 
${\bf 1}$ &  ${\bf 1}$ &  ${\bf 1}$ &  ${\bf 1}$ 
\\ \hline
\multicolumn{2}{|c}{$2-r_e$}  &  
\multicolumn{2}{c}{$2-r_\nu$}  & 
 $r_{\Theta u}$ &  $\bar{r}_{\Theta u}$ &  
\multicolumn{2}{c|}{$1-r_d$} 
\\ \hline   
\end{tabular}

\end{center}

\end{table}


\vspace{2mm}

{\bf 2.3 \ $R$ charge assignments}

In this model, the existence number of flavons is larger than that of VEV relations.
Therefore, in general, we can uniquely determine $R$ charges of flavons.
Since we  make a request to assign $R$ charges as simple as possible, we put 
the following rules: 

\noindent (i) We assign the same $R$ charge to flavons $A$ and $\bar{A}$ 
with the same VEVs, $\langle A \rangle = \langle \bar{A} \rangle$, e.g.
$$
\begin{array}{l}
R(E) = R(\bar{E})=\frac{1}{2} \equiv r_E , \\
P(\Phi_0) = R(\bar{\Phi}_0) \equiv r_0 , \\
P(\Phi_f) = R(\bar{\Phi}_f) \equiv  r_f . \\
\end{array}
\eqno(2.26)
$$
Note that we consider $R(P_u) \neq R(\bar{P}_u)$ because of
$\langle P_u \rangle \neq \langle \bar{P}_u \rangle$. 
Therefore, we obtain relations $R(S_u)= r_u + 2 \bar{r}_P - 2r_0$
and $R(\bar{S}_u)= {r}_u + 2 {r}_P - 2r_0$, separately. 
On the other hand, we take the option (2.14) for $\langle \Phi_\nu \rangle$,
which contains a cmplex parameter $a_\nu$ as seen in the next section. 
Therefore, we take $ \langle \Phi_\nu \rangle = \langle \bar{\Phi}_\nu \rangle$, so that 
$R(\Phi_\nu)=R(\bar{\Phi}_\nu) =r_\nu$.
Then, $R(\hat{Y}_f)$ is simply given by 
$$
R(\hat{Y}_f) = 2 R(\Phi_f) = 2 r_f  \ \ \ (f=e, \nu, d, u) , 
\eqno(2.27)
$$
from Eq.(2.16).

\noindent (ii) We can regard that $R$ charges of $\hat{Y}_f$ are determined 
only by those of the SU(2)$_L$ singlet fermions $f^c$.
Therefore, we simply assign 
$$
R(\ell H_u) = R(\ell H_d) = R(q H_u) = R(q H_d) = 2 .
\eqno(2.28)
$$
(Since those have different quantum number of U(1)$_Y$, 
we can distinguish those from each other.)
Then, we obtain a simple $R$ charge relation
$$
R(\hat{Y}_f) = - R(f^c) .
\eqno(2.29)
$$

For $Y_R$, we obtain
$$
R(Y_R) = 2-2R(\nu^c) = 2-2 \left( 2-R(\ell H_u) -R(\hat{Y}_\nu) \right) 
= 2 +2 R(\hat{Y}_\nu) ,
\eqno(2.30)
$$
from Eqs.(2.1) and (2.28). 
On the other hand, from Eq.(2.24), $R(Y_R)$ must be satisfied a relation 
$$
R(Y_R) = R(\Phi_u) + R(\hat{Y}_e) .
\eqno(2.31)
$$
 
If we consider $R(\hat{Y}_f)=0$, then we can attach the field $\hat{Y}_f$ on 
any term in superpotential. 
Therefore, we require $R(\hat{Y}_f) \neq 0$ for any $f=e, \nu, d, u$.
Also, we have to require  $R(\hat{Y}_f \hat{Y}_{f'}) \neq 0$ for any combination
of $f$ and $f'$. 
As a result, we have to consider that whole $R$ values of $\hat{Y}_f$  
are positive. 
Therefore, we speculate that the values of $R$ will be describe by simple integers,
so that, by way of trial, let us put  
$$
\left( R(\hat{Y}_\nu), R(\hat{Y}_u), R(\hat{Y}_e),  R(\hat{Y}_d) \right) 
= (1, 2, 3, 4).
\eqno(2.32)
$$
Then, the assignments (2.32) give
$$
\begin{array}{l}
R(Y_R) = 2+ 2 r_\nu = 2+2=4, \\ 
R(\Phi_u) + R(\hat{Y}_e) =r_u + 2r_e =1+3=4 ,
\end{array}
\eqno(2.33)
$$
so that the requirement (2.31) is satisfied. 
Note that, thus, the simple assignment of $R$, Eq.(2.32), guarantees 
the existence of the flavon interaction term (2.24), which plays a very 
important role in giving the peculiar form of neutrino Majorana mass matrix.

\vspace{2mm}

\noindent{\large\bf 3 \ Parameter fitting}

\vspace{2mm}

\noindent{\bf 3.1 \ How many parameters?}

We summarize our mass matrices $M_f$ as follows:
$$
M_e  = [\Phi_0 ( {\bf 1} + 
a_e X_3 )\Phi_0 ]^2 +\xi_e {\bf 1}\ \ \ \ (a_e=0, \, \xi_e=0),
\eqno(3.1)
$$
$$
M_D = [\Phi_0 ( {\bf 1} + 
a_\nu e^{i\alpha_\nu} X_3) \Phi_0 ]^2 +\xi_\nu {\bf 1},
\eqno(3.2) 
$$
$$
M_u = P_u \left([\Phi_0  \left( {\bf 1} + 
a_u  X_3 \right) \Phi_0]^2  +\xi_u {\bf 1} \right)P_u^*,
 \eqno(3.3)
$$
$$
M_d = \left[\Phi_0  \left( {\bf 1} + 
a_d  X_3  \right)  \Phi_0  + \xi'_d {\bf 1} \right]^2 ,
\eqno(3.4)
$$
$$
M_\nu = M_D Y_R^{-1} M_D , \ \ \ \ 
Y_R = Y_e \Phi_u  + \Phi_u Y_e . 
\eqno(3.5)
$$
Here, for convenience, we have dropped 
the notations ``$\langle$" and ``$\rangle$". 
Since we are interested only in the mass ratios and mixings, 
we use dimensionless expressions
$\Phi_0 = {\rm diag}(x_1, x_2, x_3)$ (with $x_1^2+x_2^2+x_3^2=1$),  
$P_u= {\rm diag} (e^{i\phi_1}, e^{i\phi_2},1)$, 
and $E={\bf 1}={\rm diag}(1,1,1)$. 
Therefore, the parameters $a_e$, $a_\nu$, $\cdots$ are re-defined 
by Eqs.(3.1)-(3.5).

Meanwhile, we require ``economy of the number of parameters".
Namely, we neglect parameters which play no essential roles in numerical 
fitting to the mixings and mass ratios as far as possible. 
In the present model, we assume that the parameters $a_e$, $a_u$ 
and $a_d$ are real, 
while $a_\nu$ is complex. So that 
we have denoted the parameter $a_\nu$ as $a_\nu e^{i\alpha_\nu}$ in Eq.(3.2). 
We also assume that the parameters $\xi_f$\,\,($f=e, u,$ and $\nu$) and $\xi'_d$ are real.
We consider that the charged lepton sector is the most fundamental 
flavor scheme, and the charged lepton mass matrix should take 
the most simple form. 
Therefore, we assume $a_e =0$ and $\xi_e=0$ in Eq.(3.1).
Then, the parameter values $x_1/x_2$ and $x_2/x_3$
are fixed by the charged lepton masses as 
$$
\frac{x_1}{x_2}=\left(\frac{m_e}{m_\mu}\right)^{1/4} , \ \ \ 
\frac{x_2}{x_3}=\left(\frac{m_\mu}{m_\tau}\right)^{1/4} .
\eqno(3.6)
$$ 
So we obtain
$$
(x_1, x_2, x_3) =(0.115144, 0.438873, 0.891141) ,
\eqno(3.7)
$$
where we have normalized $x_i$ as $x_1^2+x_2^2+x_3^2=1$.

Therefore, in the present model, except for the parameters 
$(x_1, x_2, x_3)$, we have 9 adjustable parameters,  
$(a_\nu, \alpha_\nu, \xi_\nu)$, $(a_u, \xi_u)$, $(a_d,  \xi'_d)$,
 and $(\phi_1, \phi_2)$   
for the 16 observable quantities (6 mass ratios in the
up-quark-, down-quark-, and neutrino-sectors, four CKM 
mixing parameters, and 4+2 PMNS mixing parameters). 
Especially, quark mass matrices $M_u$ and $M_d$ are fixed
by two parameters $(a_u, \xi_u)$ and $(a_d, \xi'_d)$, respectively.
(Note that those parameters are family-number independent 
parameters.)
Therefore, in oder to fix those parameters, we use two input
values, up-quark mass ratios $(m_u/m_c, m_c/m_t)$ and 
down-quark mass ratios $(m_d/m_s, m_s/m_b)$, respectively,
as we discuss in the next subsection 3.2.
After the parameters $(a_u, \xi_u)$ and $(a_d, \xi'_d)$
have been fixed by the observed quark mass rations, we have
five parameters $(a_\nu, \alpha_\nu, \xi_\nu)$ and $(\phi_1, \phi_2)$
as remaining free parameters. Processes for fitting those five parameters 
are listed in Table~3. 
In subsection 3.3, we discuss PMNS mixing ($\sin^2 2 \theta_{12}$, 
$\sin^2 2 \theta_{23}$, and $\sin^2 2 \theta_{13}$) 
and neutrino mass ratio 
($R_\nu \equiv {\Delta m_{21}^2}/{\Delta m_{32}^2}$)
 by adjusting three parameters $(a_\nu, \alpha_\nu, \xi_\nu)$.
Also, in subsection 3.4, we discuss four CKM mixing parameters,  
$|V_{us}|$, $|V_{cb}|$, $|V_{ub}|$ and $|V_{td}|$, by adjusting 
two parameters $(\phi_1, \phi_2)$.

Note that the purpose of the present paper is not 
to compete with other models for reducing parameter number 
in the model, but to investigate whether it is 
possible or not to fit all of the mixing parameters and 
mass ratios without using any family number dependent 
parameters when we use only the observed charged lepton 
masses as family dependent parameters.
If we pay attention only to fitting of mixing parameters, 
a model with fewer number of parameters based on quark-lepton 
complementarity \cite{q-l_compl} is rather 
excellent compared with the preset model.
(For such a recent work, see, for example,  
Ref.\cite{q-l_compl_PRD12} and references there in.)

\vspace{2mm}

\begin{table}
\caption{Process for fitting parameters. 
 $N_{parameter}$ and $N_{input}$ denote a number of free 
parameters in the model and a number of observed values which
are used as inputs in order to fix these free parameters, 
respectively. $\sum N_{\dots}$ means $\sum N_{parameter}$ or  $\sum N_{input}$
}
\vspace{2mm}
\begin{center}
\begin{tabular}{|c|cc|cc|c|} \hline
Step & Inputs & $N_{input}$ &  Parameters & $N_{parameter}$ &
 Predictions  \\ \hline
1st  &  $m_e/m_\mu$, $m_\mu/m_\tau$ & 2 & $x_1/x_2$, $x_2/x_3$ & 2 & 
--- \\
     & $m_u/m_c$, $m_c/m_t$ & 2 & $a_u$, $\xi_u$ & 2 & --- \\
     &  $m_d/m_s$, $m_s/m_b$ & 2 & $a_d$, $\xi'_d$ & 2 & --- \\
2nd  & $\sin^2 2\theta_{12}$, $\sin^2 2\theta_{23}$, $R_\nu$  & 3 
& $\xi_\nu$, $a_\nu$, $\alpha_\nu$  & 3 & 
$\sin^2 2\theta_{13}$, $\delta_{CP}^\ell$  \\
    &    &   &   &  & 2 Majorana phases, $\frac{m_{\nu 1}}{m_{\nu 2}}$,
 $\frac{m_{\nu 2}}{m_{\nu 3}}$   \\ 
3rd  & $|V_{cb}|$, $|V_{ub}|$  & 2 &   $(\phi_1, \phi_2)$ & 2 & 
$|V_{us}|$,  $|V_{td}|$, $\delta_{CP}^q$  \\
option &  $\Delta m^2_{32}$ &   & $m_{\nu 3}$ &  & 
$(m_{\nu 1}, m_{\nu 2},  m_{\nu 3})$, $\langle m \rangle$  \\
\hline
$\sum N_{\dots}$ & & 11 &  & 11 &   \\ \hline 
\end{tabular}
\end{center}
\end{table}

\vspace{2mm}

\noindent{\bf 3.2 \ Quark mass ratios}

From the observed values \cite{q-mass}
$$
r^u_{12} \equiv \sqrt{\frac{m_u}{m_c}} 
= 0.045^{+0.013}_{-0.010} , \ \ \ \ 
r^u_{23} \equiv \sqrt{\frac{m_c}{m_t}}
=0.060 \pm 0.005 ,
\eqno(3.8)
$$
at $\mu=m_Z$ \cite{q-mass}, 
we fix values of $(a_u, \xi_u)$. 
We find four solutions of $(a_u, \xi_u)$ which can give
the values (3.8). 
Only one solution
$$
(a_u, \xi_u) = (- 1.467, -0.001467) ,
\eqno(3.9)
$$
can give a reasonable prediction of the PMNS mixing
as we discuss later.  

From the observed down-quark mass ratios \cite{q-mass}
$$
r^d_{12} \equiv \frac{m_d}{m_s} = 0.053^{+0.005}_{-0.003} , \ \ \ 
r^d_{23} \equiv \frac{m_s}{m_b} = 0.019 \pm 0.006 , 
\eqno(3.10)
$$
we determine the parameters $(a_d, \xi'_d)$ as follows: 
$$
(a_d, \xi'_d) = (- 1.477, +0.0237) .
\eqno(3.11)
$$

\vspace{2mm}

\noindent{\bf 3.3 \ PMNS mixing} 

The observed values \cite{PDG12} are
$$
\begin{array}{l}
\sin^2 2\theta_{12} = 0.857 \pm 0.024, \\
\sin^2 2\theta_{23} >0.95, \\
\end{array}
\eqno(3.12)
$$
  $$
R_{\nu} \equiv \frac{\Delta m_{21}^2}{\Delta m_{32}^2}
=\frac{m_{\nu2}^2 -m_{\nu1}^2}{m_{\nu3}^2 -m_{\nu2}^2}
=\frac{(7.50\pm 0.20) \times 10^{-5}\ {\rm eV}^2}{
(2.32^{+0.12}_{-0.08}) \times 10^{-3}\ {\rm eV}^2} = 
(3.23^{+0.14}_{-0.19} ) \times 10^{-2} .
\eqno(3.13)
$$

First, we fix the parameter $\xi_\nu$ as $\xi_\nu=-0.020$ so as to
reproduce reasonable values (3.12) and (3.13).
Next, we determine the parameter values of $(a_\nu, \alpha_\nu, \xi_\nu)$ 
as follows:
$$
(a_\nu, \alpha_\nu, \xi_\nu) = ( 3.53, 8.7^\circ, -0.020).
\eqno(3.14)
$$
Here the values of $(a_\nu, \alpha_\nu, \xi_\nu)$ in Eq. (3.14) are obtained 
so as to reproduce the observed values of the PMNS mixing angles and $R_{\nu}$. 
We show the $a_\nu$ and $\alpha_\nu$ dependences of the PMNS mixing parameters 
$\sin^2 2\theta_{12}$, $\sin^2 2\theta_{23}$, $\sin^2 2\theta_{12}$, 
and $R_{\nu}$ in Fig.~1(a) and Fig.~1(b), respectively. 
It is found that $R_{\nu}$ is very sensitive to $a_\nu$. 
\vspace{10mm}

\begin{figure}[ht]
\begin{picture}(200,200)(0,0)
\includegraphics[height=.3\textheight]{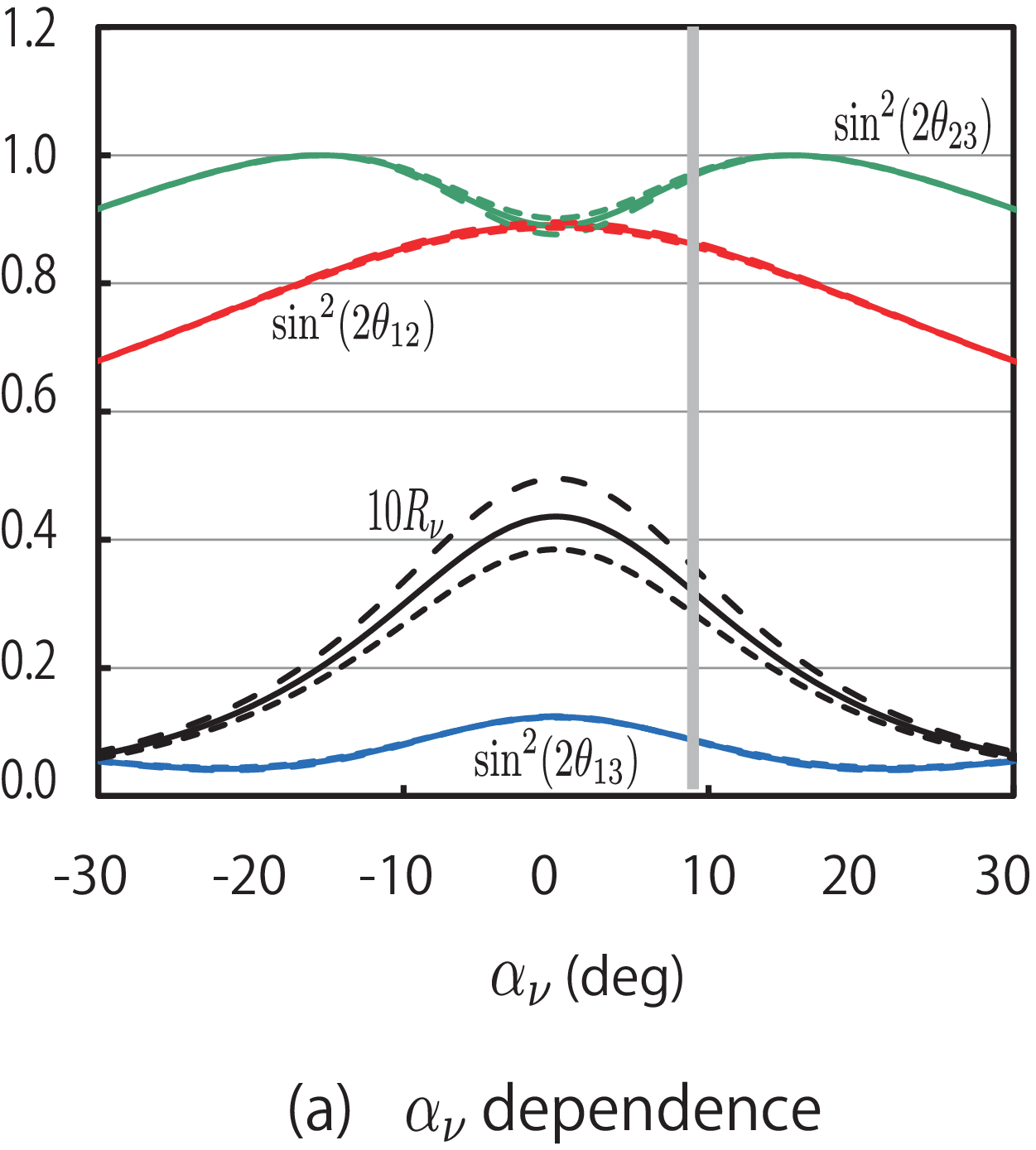}\quad \quad
\includegraphics[height=.3\textheight]{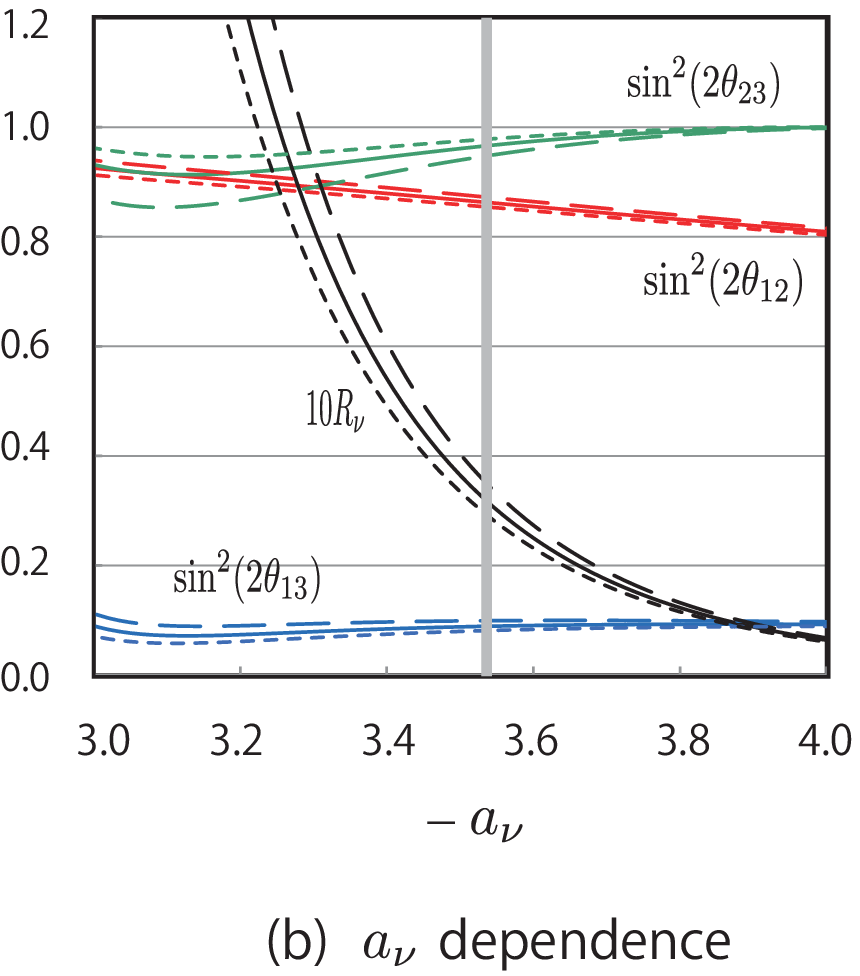}
\end{picture}  
  \caption{(a): $\alpha_\nu$ dependence of the lepton mixing parameters 
$\sin^2 2\theta_{12}$, 
$\sin^2 2\theta_{23}$, $\sin^2 2\theta_{13}$, and the neutrino 
mass squared difference ratio $R_\nu$. 
We draw curves of those as functions of $\alpha_\nu$   
for the case of $\xi_\nu=-0.20$ 
with taking $a_\nu=-3.5$ (dotted), $-3.53$ (solid), and $-3.56$ (dashed).
(b): $a_\nu$ dependence of the lepton mixing parameters $\sin^2 2\theta_{12}$, 
$\sin^2 2\theta_{23}$, $\sin^2 2\theta_{13}$, and the neutrino 
mass squared difference ratio $R_\nu$.  
We draw curves of those as functions of $a_\nu$   
for the case of $\xi_\nu=-0.20$ with taking $\alpha_\nu=7.0^\circ$ (dotted), 
$8.7^\circ$ (solid), and $10^\circ$ (dashed).
}\label{fig1}
\end{figure}

\vspace{2mm}

\noindent{\bf 3.4 \ CKM mixing}  

Next, we discuss quark sector. 
Since the parameters $(a_u, \xi_u)$ and $(a_d, \xi'_d)$
have been fixed by the observed quark mass rations, 
the CKM mixing matrix elements  $|V_{us}|$, 
$|V_{cb}|$, $|V_{ub}|$, and  $|V_{td}|$ are functions of 
the remaining two parameters 
$\phi_1$ and $\phi_2$.
In Fig.~2, we draw allowed regions in the ($\phi_1$, $\phi_2$) 
parameter plane which are obtained from the observed constraints of 
the CKM mixing matrix elements shown in Eq.~(3.15), 
with taking  $\xi_u=-0.001467$, $a_u=-1.467$, $a_d=-1.477$, and $\xi'_d=0.0237$. 
As shown in Fig.~2, all the experimental constraints on 
CKM parameters are satisfied by 
fine tuning the parameters $\phi_1$ and $\phi_2$ around 
$$
(\phi_1, \phi_2)=(21.8^\circ, -4.9^\circ ). 
\eqno(3.15)
$$
Here we have used the observed values \cite{PDG12}  
$$
\begin{array}{l}
|V_{us}|=0.22534 \pm 0.00065, \\
|V_{cb}|=0.0412^{+0.0011}_{-0.0005}, \\
|V_{ub}|=0.00351^{+0.00015}_{-0.00014}, \\
|V_{td}|=0.00867^{+0.00029}_{-0.00031}.\\
\end{array}
\eqno(3.16)
$$

\begin{figure}[ht]
\begin{picture}(200,200)(0,0)

  \includegraphics[height=.33\textheight]{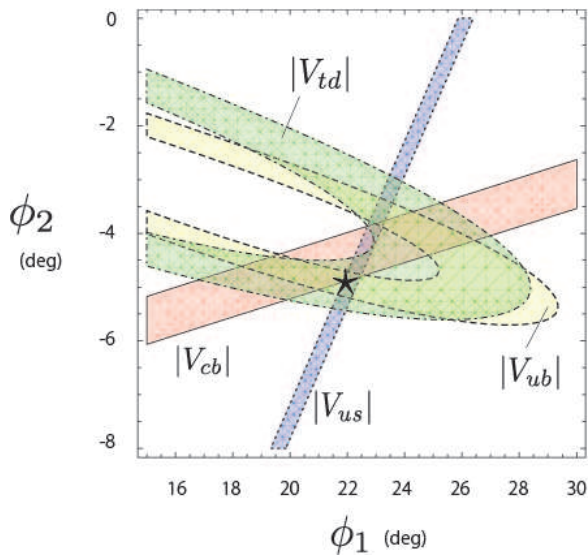}

\end{picture}  
  \caption{Allowed region in the ($\phi_1$, $\phi_2$) parameter plane obtained 
by the observed values of the CKM mixing matrix elements  $|V_{us}|$, 
$|V_{cb}|$, $|V_{ub}|$, and  $|V_{td}|$. 
We draw allowed regions obtained from the observed constraints of the CKM mixing 
matrix elements shown in Eq.~(3.15), 
with taking  $\xi_u=-0.001467$, $a_u=-1.467$, $a_d=-1.477$, and $\xi'_d=0.0237$. 
Here we take $2\sigma $ errors for all the observed values of the CKM mixing matrix elements.
We find that the parameter set arround ($\phi_1$, $\phi_2$) 
$=(21.8^\circ, -4.9^\circ)$ 
indicated by a star ($\star$) is 
consistent with all the observed values.} 
 \label{fig2}
\end{figure}

\vspace{2mm}

\noindent{\bf 3.5 \ Neutrino masses and leptonic Dirac $CP$ violating phase}  

We can predict neutrino masses, for the parameters given by (3.9) and (3.14), 
as follows

$$
m_{\nu 1} \simeq 0.00040\ {\rm eV}, \ \ m_{\nu 2} \simeq 0.00890 \ {\rm eV}, 
\ \ m_{\nu 3} \simeq 0.0501 \ {\rm eV}  ,
\eqno(3.17)
$$
by using the input value \cite{MINOS13}
$\Delta m^2_{32}\simeq 0.00241$ eV$^2$.

We also predict the effective Majorana neutrino mass \cite{Doi1981} 
$\langle m \rangle$ 
in the neutrinoless double beta decay as
$$
\langle m \rangle =\left|m_{\nu 1} (U_{e1})^2 +m_{\nu 2} 
(U_{e2})^2 +m_{\nu 3} (U_{e3})^2\right| 
\simeq 5.1 \times 10^{-3}\ {\rm eV}.
\eqno(3.18)
$$

Our model also predicts $\delta_{CP}^{\ell}= 25.7^\circ$ for the Dirac $CP$ 
violating phase in the lepton sector, which indicates relatively large  $CP$ 
violating effect in the lepton sector.
(Note that the previous model predicts 
$\delta_{CP}^{\ell}= 179^\circ$ which indicates small $CP$ 
violating effect in the lepton sector. )

\begin{table}
\caption{Predicted values vs. observed values. 
}

\vspace*{2mm}
\hspace*{-6mm}
\begin{tabular}{|c|ccccccccc|} \hline
  & $|V_{us}|$ & $|V_{cb}|$ & $|V_{ub}|$ & $|V_{td}|$ & 
$\delta^q_{CP}$ &  $r^u_{12}$ & $r^u_{23}$ & $r^d_{12}$ & $r^d_{23}$ 
 \\ \hline 
Pred &$0.2225$ & $0.0430$ & $0.00405$ & $0.00800$ & $55.8^\circ$ & 
$0.0416$ & $0.0627$ & $0.0492$ & $0.0192$ 
 \\
Obs & $0.22534$ & $0.0412$ &  $0.00351$  & $0.00867$  & $68^\circ$ &
$0.045$ & $0.060$ & $0.053$  & $0.019$  
  \\ 
    &  $ \pm 0.00065$ &  $ {}^{+0.0011}_{-0.0005}$ & $ {}^{+0.00015}_{-0.00014}$ & 
 ${}^{+0.00029}_{-0.00031}$ &
 ${}^{+10^\circ}_{-11^\circ}$ &
${}^{+0.013}_{-0.010}$ & $ \pm 0.005$ & $^{+0.005}_{-0.003}$ &
${}^{+0.006}_{-0.006}$ 
 \\ \hline
   & $\sin^2 2\theta_{12}$ & $\sin^2 2\theta_{23}$ & $\sin^2 2\theta_{13}$ & 
 $R_{\nu}\ [10^{-2}]$ &  
$\delta^\ell_{CP}$ & $m_{\nu 1}\ [{\rm eV}]$ & $m_{\nu 2}\ [{\rm eV}]$ & 
$m_{\nu 3}\ [{\rm eV}]$ & $\langle m \rangle \ [{\rm eV}]$ \\ \hline
 Pred & $0.863$ & $0.965$ & $0.089$ &  $3.25$ & $25.7^\circ$ &
 $0.00040$ & $0.00890$ & $0.0501$ & $0.00514$ \\
Obs & $0.857$   & $ >0.95$ & $0.095$ &   $3.23 $    & -
  &  -  &  -  &  -  &  $<\mathrm{O}(10^{-1})$   \\ 
    & $ \pm 0.024$ &   & $\pm0.010$ & ${}^{+0.14}_{-0.19} $  &  &
   &    &    &    \\ \hline 
\end{tabular}
\end{table}

\vspace{3mm}

\noindent{\large\bf 4 \ Concluding remarks}

We have tried to describe quark and lepton mass matrices
by using only the observed values of charged lepton masses 
$(m_e, m_\mu, m_\tau)$ as input parameters with family-number dependent values. 
Thereby, we have investigated  
whether we can describe all other observed mass spectra 
(quark and neutrino mass spectra) and mixings (CKM and 
PMNS mixings) without using any other family-number 
dependent parameters. 
In conclusion, as seen in Sec.3, we have obtained reasonable 
results.
Our predicted values are listed in Table~4.

However, we have been still obliged to bring a family-number
dependent VEV matrix $P_u$ given in Eq.(2.3). 
When we consider that our aim has been completed except for
only $P_u$, and that it appears only in the quark sector, 
there is a possibility that the origin of the matrix
form $P_u$ is not due to a VEV form of a flavon $P_u$,
but it may be due to another origin, for example, a dynamical 
origin such as QCD effects, and so on. 
This is an open question at present. 

In the present revised version of yukawaon model, the following
points are worthy of note: 

\noindent (i) 
We have been able to describe the VEV matrices of the yukawaons with 
the unified forms $\hat{Y}_f = \Phi_f \bar{\Phi}_f$.

\noindent (ii) 
Especially, we have adopted
a bilinear form for charged lepton mass matrix, 
$\hat{Y}_e = \Phi_e \bar{\Phi}_e$. 
It is  for the first time to succeed in giving a large value 
$\sin^2 2\theta_{13}\sim 0.09$ 
without taking a non-diagonal form of $\hat{Y}_e$. 
By this model-change, the charged lepton mass formula (1.4)
has again become possible to understand from  
${\rm Tr}[\Phi_e\Phi_e] = \frac{2}{3} {\rm Tr}[\Phi_e] {\rm Tr}[\Phi_e]$
although we did not discuss the relation (1.4) in the present paper.

\noindent (iii)
The VEV relation of $Y_R$ to $\Phi_u$ and $\hat{Y}_e$, Eq.(2.25), is 
ad hoc assumption in the past models \cite{yukawaon_models,K-N_PRD13}. 
(The $R$-charges have been assigned so that the
ad hoc relation $R(Y_R) =R(\Phi_u) + R(\hat{Y}_e)$ may be satisfied.)
In the present model, we have demonstrated that a simple 
$R$ charge assignment (2.32) guarantees the relation (2.25). 
At present, the meaning of the assignment (2.32) is unclear, investigation of which is left to our future task.

\noindent (iv)
In the present model, we have predicted   
the $CP$ violating phase in the lepton sector as  
 $\delta_{CP}^{\ell} \simeq 26^\circ$, which is sufficiently large to observe $CP$ 
violation effects in future experiments. 
(In the previous model \cite{K-N_PRD13}, a predicted value of $\delta_{CP}^{\ell}$ 
was $\delta_{CP}^{\ell} \simeq 179^\circ$, which was invisibly small.)
The origin of the $CP$ violation is in the phase factor $\alpha_\nu$
in the Dirac neutrino mass matrix (3.2). 
Note that we have taken $\alpha_f =0$ ($f=e, u,d$) for economy of 
the parameters.  
However, we have been obliged to accept $\alpha_\nu \neq 0$ 
in order to fit the observed value of $\sin^2 2\theta_{13}$.

We still have some open questions as follows:

\noindent(a)
Compared with the previous model \cite{K-N_PRD13},
number of free parameters is not so reduced in the present 
yukawaon model.
As emphasized in Sec.1, the purpose of the present paper 
is not to build a model with economized parameters.
In the present yukawaon model, the VEV relations among 
flavons have been given by universal forms compared with 
those in the past yukawaon models \cite{yukawaon_models}. 
Some of the parameters in the past yukawaon models
have been eliminated, but, instead, terms
which shift VEV matrices of yukawaons by unit matrices 
$\xi_f {\bf 1}$  (or $\xi^{\prime}_f {\bf 1}$) have been 
newly added in the present model. 
This means that the present model cannot give predictions
as far as the mass ratios are concerned, and it is nothing but 
that two parameters ( $a_f$ and $\xi_f$) or ($a_d$ and $\xi'_d$) are fixed by the 
two observed mass ratios. 
Therefore, in the present model, only mixings can be predicted 
as far as quark sector is concerned. 

\noindent(b) 
In spite of our aim to describe 
whole of quark and lepton masses and mixings by using
only the observed charged lepton masses as input parameters
with hierarchical values, we again need family-number 
dependent parameters $(\phi_1, \phi_2)$ in the description 
of the CKM mixing. 
Also the origin of $CP$ violation in the quark sector is 
in the phase matrix
$P_u$, i.e. the phase parameters $(\phi_1, \phi_2)$. 
[Note that in the lepton sector the origin of $\delta_{CP}^\ell \neq 0$
is $\alpha_\nu \neq 0$ which is inevitably required in order to get 
reasonable fitting of the PMNS mixing angles and the neutrino mass ratio $R_\nu$.]
Namely, we have different origins of $CP$ violations between lepton and 
quark sectors. 
This is still unsatisfactory to us.
The phase matrix $P_u$ has family-number dependent parameters 
$(\phi_1, \phi_2)$, so that such parameters should be eliminated 
in the final goal of the yukawaon model. 
We consider that, in a yukawaon model at the final goal, 
the $CP$ violation in the quark sector, too,  should be brought by 
family-number independent parameters $\alpha_u$, $\alpha_d$, and so on.

By success of the present major improvement of the yukawaon model,
it seems that we are considerably close to the ideal stage that 
all hierarchical structures of quarks and leptons can be 
understood only from the family-number dependent parameter 
values $(m_e, m_\mu,m_\tau)$.  
However, at present, we have many flavons and free parameters.
Our next task is to economize numbers of those flavons and 
free parameters.

\vspace{3mm}

\vspace{2mm}
%

\end{document}